\documentclass[usegraphicx,useAMS]{mn2e}
\usepackage{rotating,graphicx,graphics,subfigure,epsf,amsmath,amssymb}

\newcommand{\hi}{H\,{\sc i}}

\newcommand{\degree}{$^{\circ}$}
\newcommand{\kms}{km s$^{-1}$}
\newcommand{\etal}{{\it et al.}}
\newcommand{\rah}{$^h$}
\newcommand{\ram}{$^m$}
\newcommand{\ras}{$^s$}

\title[{\it Herschel} observations of Cen A dust clouds] {{\it Herschel} observations of Cen A: stellar heating of two extragalactic dust clouds}
\author[Auld \etal]{R. Auld$^1$\footnote{email: robbie.auld@astro.cf.ac.uk}, M. W. L. Smith$^1$, G. Bendo$^2$, M. Pohlen$^1$, C. Wilson$^3$ , H. Gomez$^1$, \newauthor L. Cortese$^4$, R. Morganti$^{5,6}$, M. Baes$^7$, A. Boselli$^8$, A. Cooray$^{9,10}$, J. I. Davies$^1$, \newauthor S. Eales$^1$, D. Elbaz$^{11}$, M. Galametz$^{12}$, K. Isaak$^{13}$, T. Oosterloo$^{5,6}$,  M. Page$^{14}$, \newauthor E. Rigby$^{15}$, L. Spinoglio$^{16}$, C. Struve$^{5,6}$\\
$^1$School of Physics and Astronomy, Cardiff University, Queens Buildings, The Parade, Cardiff CF24 3AA, United Kingdom\\
$^2$Jodrell Bank Centre for Astrophysics, School of Physics and Astronomy University of Manchester, Oxford Road, Manchester M13 9PL, UK\\
$^3$Department of Physics \& Astronomy, McMaster University, Hamilton, Ontario L8S 4M1, Canada\\
$^4$European Southern Observatory, Karl-Schwarzschild Str. 2, 85748 Garching bei Muenchen, Germany\\
$^5$Netherlands Institute for Radio Astronomy, Postbus 2, 7990 AA, Dwingeloo, The Netherlands\\
$^6$Kapteyn Astronomical Institute, University of Groningen, P.O. Box 800, 9700 AV Groningen, The Netherlands\\
$^7$Sterrenkundig Observatorium, Universiteit Gent, Belgium\\
$^8$Laboratoire d\u2019Astrophysique de Marseille, France\\
$^9$Dept. of Physics \& Astronomy, University of California, Irvine, CA 92697, USA\\
$^{10}$ California Institute of Technology, 1200 E. California Blvd., Pasadena, CA 91125, USA\\
$^{11}$Laboratoire AIM-Paris-Saclay, CEA/DSM/Irfu - CNRS - Universit\'e
Paris Diderot, CE-Saclay, pt courrier 131, F-91191 Gif-sur-Yvette,
France\\
$^{12}$Institute of Astronomy, University of Cambridge, Madingley Road, Cambridge CB3 0HA, UK\\
$^{13}$European Space \& Technology Centre (ESTEC), P.O.Box 299, 2200 AG Noordwijk, The Netherlands\\
$^{14}$UCL, Mullard Space Science Laboratory, Holmbury St. Mary, Dorking, Surrey RH5 6NT, UK\\ 
$^{15}$School of Physics and Astronomy, University of Nottingham, University Park, Nottingham NG7 2RD, UK\\
$^{16}$IFSI Roma, Via del Fosso del Cavaliere, 100 00133 Roma, Italy}
\begin{document}
\setlength{\paperheight}{297mm}
\setlength{\paperwidth}{210mm}

\date{Accepted 11 September 2011.
      Received 24 August 2011;
      in original form 11 May 2011}
\pagerange{\pageref{firstpage}--\pageref{lastpage}} \pubyear{2007}

\maketitle

\label{firstpage}

\begin{abstract}
We present the first results of a multi-wavelength survey, incoporating {\it Herschel-SPIRE}, {\it Spitzer}, {\it GALEX} and {\it ATCA} observations, of a 1\degree$\times$1\degree\ field centred on Centaurus A. As well as detecting the inner lobes of the active galactic nucleus (AGN) jet and counterjet, we have found two clouds, bright at sub-mm wavelengths, $\sim$15 kpc from the centre of Cen A that are co-aligned with the jets. Flux measurements at {\it Herschel} wavelengths have proved vital in constraining fits to the Spectral Energy Distributions (SEDs). The clouds are well fit by a single-temperature, modified blackbody spectrum ($\beta = 2$) indicating that we are looking at two cold dust clouds on the outskirts of Cen A. The temperature and masses of the clouds are: T$_{north} = 12.6^{+1.1}_{-1.2}$ K, T$_{south} = 15.1^{+1.7}_{-1.6}$ K; $\rm{log} (M_{north} / M_{\odot}) = 5.8^{+0.2}_{-0.2}$, $\rm{log} (M_{south} / M_{\odot}) = 5.6^{+0.2}_{-0.2}$ and the gas-dust ratio for both clouds is $\sim$100. The measured values for the northern dust cloud are consistent with previous measurements from {\it ISO} while the southern cloud is a new sub-mm detection. The two dust clouds are located at the termini of the partial \hi\ ring that surrounds Cen A which is also where the gas column density peaks. The {\it Herschel} survey encompasses the partial \hi\ ring yet we find no evidence of dust emission in any other part of the ring. Assuming that the gas-dust ratio is the same in the rest of the ring, dust mass upper limits in the \hi\ ring are consistent with low column density dust being present but falling below the {\it SPIRE} detection limit. We have discussed the origin of these clouds and various possible heating mechanisms. The observations favour a scenario in which the gas and dust were once part of a late-type galaxy which has since merged with Cen A. The dominant heating mechanism which adequately explains the observed temperatures in both clouds is heating from the evolved stellar population within Cen A. 
\end{abstract}

\section{Introduction}
\subsection{Cen A and its environment}
Centaurus A (NGC 5128) is the dominant, elliptical galaxy at the heart of the Centaurus group, 3.6 Mpc away (Chattopadhyay \etal,  2009). It exhibits a  complex variety of gas, dust and stellar structures, observable throughout the electromagetic spectrum, on scales ranging from sub-pc to kpc scales. 

Optical broadband images are dominated by an aging stellar population (Soria, 1996), arranged in a system of shells (Peng \etal\ 2002), extending out to $\sim$15 kpc. Bisecting the galaxy along its minor axis is a prominent dust lane, first observed by John Herschel (1847). The optical shells and the presence of a large amount of dust are unusual properties for an elliptical galaxy and are indicative of a past merger with a late-type galaxy, rich in dust and gas (Baade \& Minkowski, 1954; Malin, 1983).

Optical spectroscopy of the interstellar medium has revealed star forming regions and filaments of highly excited gas (Blanco \etal, 1975; Peterson \etal, 1975; Morganti \etal, 1991; Morganti \etal, 1992; Rejkuba \etal, 2002). The high ionisation states have been interpreted as either due to a beamed ionising source emanating from the nucleus (Morganti \etal, 1991) or shocks being driven into the Interstellar Medium (ISM, Bicknell, 1991; Sutherland, Bicknell \& Dopita, 1993; Fragile, \etal, 2004).

At radio wavelengths, spectral line observations have detected atomic, molecular and ionised gas in the dusty disk, while wide-field observations in the near vicinity of Cen A have revealed gas in a huge broken ring surrounding the galaxy at a projected distance of $\sim$15 kpc. The ring contains substantial amounts of atomic (Shiminovich \etal, 1994; Oosterloo \& Morganti, 2005; Struve, 2010) and molecular gas (Charmandaris \etal, 2000) and the velocity structure varies smoothly over the extent of the ring, consistent with rotation about the centre of Cen A. An exception to this trend in velocity can be found at the northern terminus, where Oosterloo \& Morganti (2005) have discovered highly turbulent atomic gas. 

Radio continuum images reveal huge lobes extending north and south from the galaxy over $\sim$10\degree. The radio emission is dominated by synchotron emission from a jet originating in the AGN of Cen A and this is driving plasma at relativistic speeds into the surrounding intergalactic medium up to distances of 100s kpc. This ongoing interaction results in shock heating of the ISM, clearly visible as extended, knot-like features in X-ray images (Kraft \etal, 2003, 2007, 2009; Croston, 2009). Although the northern and southern jets exhibit symmetry at large scales, there are crucial differences between them. Both have an inner lobe extending to $\sim$5 kpc and an outer lobe extending to 100s kpc. The difference lies in the existence of a northern middle lobe (NML), which is connected to and powered by the central engine via a large-scale jet (Junkes \etal, 1993; Morganti \etal, 1999; Kraft \etal, 2009; Hardcastle \etal, 2009). In contrast, the southern jet appears to terminate at the boundary of the inner southern lobe where it is driving a strong shock into the ISM (Kraft, 2003, 2007; Croston, 2009).

One aspect of AGN that has come under close scrutiny in recent years is their supposed ability to regulate star formation in the galactic environment by injecting energy into the interstellar medium (ISM). This influx of energy prevents atomic gas from cooling and condensing into molecular clouds from which stars would otherwise form (e.g. Silk \& Rees, 1998). This theoretical mechanism has been invoked to try and recover the apparent discrepancy between the observed population of large, blue galaxies and those predicted by cosmological simulations (Benson \etal, 2004; Begelman \& Ruszkowski, 2005; Binney, 2005). Yet it has also been demonstrated that AGN have the ability to {\it create} stars via the interaction of the jet with gas in the near vicinity of the AGN host galaxy (Rees 1989; van Breugel \etal, 2004 and references therein). In terms of affecting a galaxy's evolution, these two phenomena seem to be at odds with one another and clearly the net effect on the galaxy's star formation rate (SFR) will depend on the relative efficiencies of these two processes. Cen A has been the focus of much debate in this respect, with numerous lines of evidence in support of jet-induced star formation taking place in Cen A (Blanco \etal, 1975; Graham, 1998; Mould \etal, 2000; Rejkuba \etal, 2002; Morganti \& Oosterloo, 2005). 

Observations of Cen A have recently been extended into the sub-mm with the detection of a cold dust cloud $\sim 15$ kpc north of the nucleus (Stickel \etal, 2004). Using the ISO observations at 90, 150 and 200 \micron\ Stickel \etal\ were able to model a SED applying a single-temperature modified blackbody spectrum ($\beta=2$), which is typical of thermally-emitting cold dust. The resulting fit implied the dust cloud has a very low temperature (13 $\leq$ T $\leq$ 15.5 K) and at the distance of Cen A, an inferred dust mass range of 4.5 $\leq {\rm log} (M_{dust}) \leq 4.9$. Unfortunately the peak of the SED for very cold dust ($<15$ K) occurs longwards of 200 \micron. With only two detections and an upper limit at 90 \micron, the fit was constrained to the rising side of the SED. One of the niches that {\it Herschel} fills, that we exploit in these observations, is that its particular wavelength coverage (70\micron -- 500\micron) enables observations of both sides of the peak, thus allowing for stricter constraints on the SED fit to the dust thermal emission.

In their discussion of the origin of the cloud, Stickel \etal\ combined their dust mass with the gas mass estimated from Schiminovich \etal\  (1994) and Charmandaris \etal\  (2000) and found that the gas-dust ratio for this cloud is approximately 300. This value falls within the range exhibited by late-type galaxies (Sodroski \etal, 1994; Stickel \etal, 2000). The ratio of atomic gas to molecular gas is close to unity, which is also consistent with values observed in spirals (Charmandaris \etal, 2000). This points towards the late-type galaxy, with which Cen A is believed to have recently merged and is responsible for the dust and gas in the centre of Cen A, as also being the source of the gas and dust observed in the external ring.

Stickel \etal\ discuss at length the heating source of this cloud. In particular they focus on the ability of stellar light from Cen A to heat the cloud directly. Temi \etal\  (2003) developed a model for predicting the temperature of different types of dust grains at various distances from the centre of early-type galaxies. The model accounts not only for reprocessed stellar emission, but also for electron-grain interactions one would expect in a hot ISM, typical of early-type galaxies. One of the main results was that grains of different sizes are dominated by different heating mechanisms. Small grains ($< 0.1 $\micron) are predominantly heated by electron-grain collisions in the hot gas, while large grains ($\sim$1 \micron) are heated by the stellar emission of the evolved stellar population.

This model was then used by Stickel \etal\ to demonstrate that the temperature of the dust in the northern cloud is fully consistent with the temperature predicted from the Temi \etal\ model, applied to NGC 4636. Since NGC 4636 is similar in morphology, size and mass to Cen A (Israel, 1998), they concluded that the FIR emission is mostly due to dust heated by the ambient starlight from Cen A without the need for heating by other sources such as embedded star formation.

Since their survey field did not encompass the whole \hi\ ring, they were unable to probe the existence of dust elsewhere in the gas ring. Also the resolution of ISO was not high enough to allow them to comment on the apparent alignment between the dust and the AGN jet. The superior sensitivity, mapping speed and resolution of {\it Herschel} present us with the opportunity to map the entire \hi\ ring in the sub-mm to unprecedented levels of sensitivity and detail.  For more information on Cen A, Israel (1998) provides a comprehensive review, while Morganti (2010) complements this picture with discoveries that have been made in the interim.

\subsection{Very Nearby Galaxy Survey}
The Very Nearby Galaxy Survey (VNGS, P.I.: C. Wilson\footnote{email:  wilson@physics.mcmaster.ca}) is a {\it Herschel} Guaranteed-Time Key Project focussing on twelve galaxies within 25 Mpc and the archetypal starburst galaxy, Arp 220. The galaxies show a diverse range of masses and properties, from low-mass, late-type galaxies such as NGC 2403 to the large elliptical Cen A, with its energetic active nucleus.

The choice of galaxies was also driven by the wealth of ancilliary data that are available. The design of the survey incoporates both photometry (SPIRE, PACS) and spectrometry (SPIRE-FTS, PACS spectrometer). With such a great breadth of multi-wavelength data the survey aims to gain a detailed understanding of the processes that regulate the ISM, and how these processes vary with the environment within different galaxies. The detailed study of these resolved galaxies will not only act as a benchmark for studies of more distant galaxies, but will also bridge the gap between surveys of distant objects and the extensive Galactic surveys which have superior physical resolution but by their nature are limited to observations of one galaxy. Recent results from the VNGS include determining the nature of dust heating in M81 (Bendo \etal, 2010), attempting to separate Galactic cirrus emission from extragalactic dust emission  in M81 (Davies \etal, 2010) and studying the dust distribution and heating mechanisms in and around the starburst galaxy, M82 (Roussel, 2010; Panuzzo, 2010). While this paper concentrates on the external dust around Cen A, a detailed analysis of the dust within Cen A will be discussed in Parkin \etal\  (2011, in prep). 

In Section \ref{obsdr} we detail the {\it Herschel} observations and the data reduction techniques that were employed. We also give a brief summary of each of the ancilliary datasets that were used in the analysis. In Section \ref{results} we compare multi-wavelength images of the clouds as seen in {\it SPIRE}, {\it Spitzer}, {\it GALEX} and {\it ATCA} and show the results of SED fitting to the two dust clouds. In Section \ref{discussion} we discuss the origin of the dust clouds and the heating machanisms which are powering the sub-mm emission. Finally in Section \ref{conclusion} we summarise our findings.

\section{Observations and Data Reduction}
\label{obsdr}
\subsection{Sub-mm observations}
The {\it Herschel} observations of Cen A consist of a 1\degree$\times$1\degree\  area centred on Cen A itself and mapped by both SPIRE (250, 350 \& 500 \micron) and PACS (70 \& 160 \micron) using Large Scan-map mode. At the time of writing, the PACS data reduction pipeline was optimised for the analysis of the central dust lane in Cen A, rather than the analysis of the dust clouds themselves,  hence the regions in the final images surrounding the clouds suffered from degraded surface brightness sensitivity. For these wavelengths MIPS maps (described in Section \ref{spitzer}) proved to have lower surface brightness limits and so were used in place of PACS.

For the SPIRE observations, the telescope was driven at a scan rate of 30\arcsec s$^{-1}$ and the observations repeated five times. The SPIRE photometer (Griffin \etal, 2010) data were processed up to Level 1 (i.e. calibrated bolometer timelines) with a custom made pipeline adapted from the official pipeline \footnote{See Griffin (2008) or Dowell (2010) for a more detailed description of the offical pipeline and a list of the individual modules.}. This Jython script was run in the Herschel Interactive Processing Environment (Ott, 2010). The purpose of the pipeline is to remove all instrumental artefacts such as glitches, finite bolometer response time, electronic filtering and thermal drift, as well as applying astrometry and flux calibration. 

The main difference between our pipeline and the standard one is that we did not run the default {\it temperatureDriftCorrection} and the residual, median baseline subtraction. Instead we use a custom method called the {\it BRIght Galaxy ADaptive Element } ({\bf BriGAdE}) to remove the temperature drift. No further baseline subtraction was necessary to bring the bolometer baselines to a common level.
 
BriGAdE uses the information from the thermistors in each array and directly fits the thermistor data to the entire bolometer timeline (including data where the spacecraft was slewing between scans). If both thermistors exhibited instantaneous `jumps' (an artefact where there is a sudden DC offset in the timelines) these are either corrected or the comparison switched to the slightly less sensitive Dark Pixels of the individual array. This approach is hampered slightly by the signal of bright sources in the bolometer timelines. To suppress their influence, they are automatically removed from the fitting process along with samples affected by other artefacts (`jumps' and glitches) in the timelines. When a choice of thermistor is available (i.e. for the 250 \& 500 \micron\ arrays), the one providing the best fit is used to subtract a scaled version from the bolometer timelines. This method improves the baseline subtraction significantly, especially in cases where there are large or rapid temperature variations during the observations.

For the final maps, we used the inbuilt na\"ive mapper of HIPE. The final maps have pixel sizes of 6\arcsec,8\arcsec \& 12\arcsec at 250, 350 \& 500 \micron\ respectively and instrument-associated noise levels of 5, 5 \& 6 mJy/beam. 

\subsection{Mid- \& far-infrared observations}
\label{spitzer}
Cen A was observed with the Spitzer observatory using the IRAC and MIPS instruments. We used 24, 70, and 160 \micron\  MIPS raw data from astronomical observational requests (AORs) 4940288, 4940544, and 4940800 retrieved from the NASA/IPAC Infrared Science Archive to produce the MIPS images.  The data from AOR 4940288 were taken at the medium scan rate and cover regions of $\sim$20$\times$50\arcmin\ that include the entire optical disc in each wave band.  The data from AORs 4940544 and 4940800 were taken at the fast scan rate and cover regions of $\sim$6$\times$115\arcmin\ that only cover the jets and the centre of the galaxy.  The resulting 160 \micron\  maps from these fast scan data contains gaps, although the 24 and 70 \micron\  data completely cover the field of view.

The data were processed using the MIPS Data Analysis Tools (Gordon \etal, 2005) along with additional processing steps decribed by Bendo \etal\  (2010) and Bendo \etal   (2011, in prep.).  The additional processing steps include an additional flatfielding step for the 24 \micron\  data; the removal of latent images from the 24 \micron\  data; background subtraction in all three wave bands (including zodiacal light subtraction in the 24 \micron\  band and short term drift removal in the 70 and 160 \micron\  bands); and a nonlinearity correction applied to the 70 \micron\  data.  For the background subtraction to be performed correctly, it is important to identify regions that are not part of the background; we explicitly identified the optical disk and the structures seen in the jets in the SPIRE images as regions to be excluded when calculating background levels.

The FWHM of the three MIPS beams are 6 arcsec at 24 \micron, 18\arcsec\  at 70 \micron, and 38\arcsec\  at 160\micron.  The rms background levels in the final images are 0.0621, 0.463, and 0.553 MJy sr$^{-1}$ at 24, 70, and 160 \micron, respectively, and the calibration uncertainties are 4\% at 24 \micron\  (Engelbracht \etal, 2007), 10\% at 70 \micron\  (Gordon \etal, 2007), and 12\% at 160 \micron\  (Stansberry \etal\  2007).

IRAC basic calibrated data frames at 3.6 \& 8 \micron\ from AOR 4939008 were taken straight from the NASA/IPAC Infrared Science Archive. The MOPEX software (v18.7.0) was used to interpolate and co-add the individual frames into one large mosaic. The 3.6 \& 8 \micron\  FWHMs are 1.7" \& 2.0" respectively and the corresponding noise levels in the final mosaicked images are 0.025 and 0.06 MJy sr$^{-1}$.

As noted in Helou \etal\ (2004) the 8 \micron\  band contains contributions from both stellar emission and PAH emission. We adopted their technique to isolate the PAH emission: foreground stars were identified by their 3.6--8 \micron\ colours - all unresolved objects with 3.6/8 surface brightness $\ga 4.5$  were masked; the 3.6 \& 8 \micron\  images were then multiplied by the aperture correction factors for extended sources (0.944 and 0.737 at 3.6 \& 8 \micron\ respectively); finally the stellar contribution was subtracted from the 8 \micron\ band using the formula:

\begin{equation}
I_{\nu} {\rm (PAH)} = I_{\nu} {\rm (8 \micron)} - 0.232 I_{\nu} {\rm (3.6 \micron)}
\end{equation}

The final PAH map was then smoothed to the FWHM of the {\it SPIRE} 250 \micron\ beam to enhance the extended emission and to allow comparison with the 250 \micron\ map.

\subsection{Radio observations}
Deep 20-cm continuum and 21-cm spectral line observations of Cen A taken with the Australian Telescope Compact Array (ATCA) already exist in the literature (20cm continuum: Morganti \etal, 1999; \hi: Oosterloo \& Morganti, 2005, Struve \etal, 2010). The authors were kind enough to make the data available for our use and we only summarise the observations here.

The continuum observations were conducted with the ATCA in multiple configurations to maximise the brightness temperature sensitivity while retaining high angular resolution. The on-source time was 12 hours in each configuration. The final size of the cleaned beam in the restored image we have used in our analysis is 56\arcsec$\times$26\arcsec\  (PA = 2\degree) and the rms noise is 6.7 mJy/beam.

The \hi\  observations were also conducted using the ATCA using a combination of two configurations. Due to the strength of the continuum source in Cen A, they were restricted in the use of short baselines, which imposed a high-pass filter on the final reduced images. This resulted in all emission on scales larger than 7\arcmin\ being filtered out. This however does not affect our analysis. The size and shape of the restored beam in the final datacube was 29\arcsec$\times$19.5\arcsec\  (PA = -4.7\degree). A moment map was produced from the datacube with a 3-$\sigma$ detection limit of 1.0$\times 10^{20} {\rm cm}^{-2}$ over a velocity width of 13.2\kms. For further details the reader is referred to Morganti \etal\  (1999) and  Oosterloo \& Morganti (2005).

\subsection{Ultraviolet observations}
Cen A was observed by GALEX (Martin \etal, 2005) as part of the Nearby Galaxy Survey (Gil de Paz, 2004). The observations were carried out simultaneously in far-ultraviolet ($FUV$: $\lambda_{eff} = 1539$ \AA, $\Delta\lambda = 442$ \AA) and near-ultraviolet ($NUV$: $\lambda_{eff} = 2316$ \AA, $\Delta\lambda = 1060$ \AA). The images used here were taken from the GALEX GR6 data release and have total exposure times of 20101 and 30460 s in FUV and NUV. 

The final data products consist of circular images $\sim$1.2\degree\ in diameter with a resolution of 4.5\arcsec\  ($FUV$) and 6\arcsec\  ($NUV$). The surface brightness limits for the FUV and NUV bands are, 30.6 and 29.5 mag arcsec$^{-2}$ respectively. The reader is referred to Morrissey \etal\  (2007) for details of the instrument, the obervations and data reduction. 

\section{Results}
\label{results}
\subsection{{\it SPIRE} images}
\label{spireimages}
\begin{figure*}
\subfigure{
	\includegraphics[angle=0,width=1.0\textwidth]{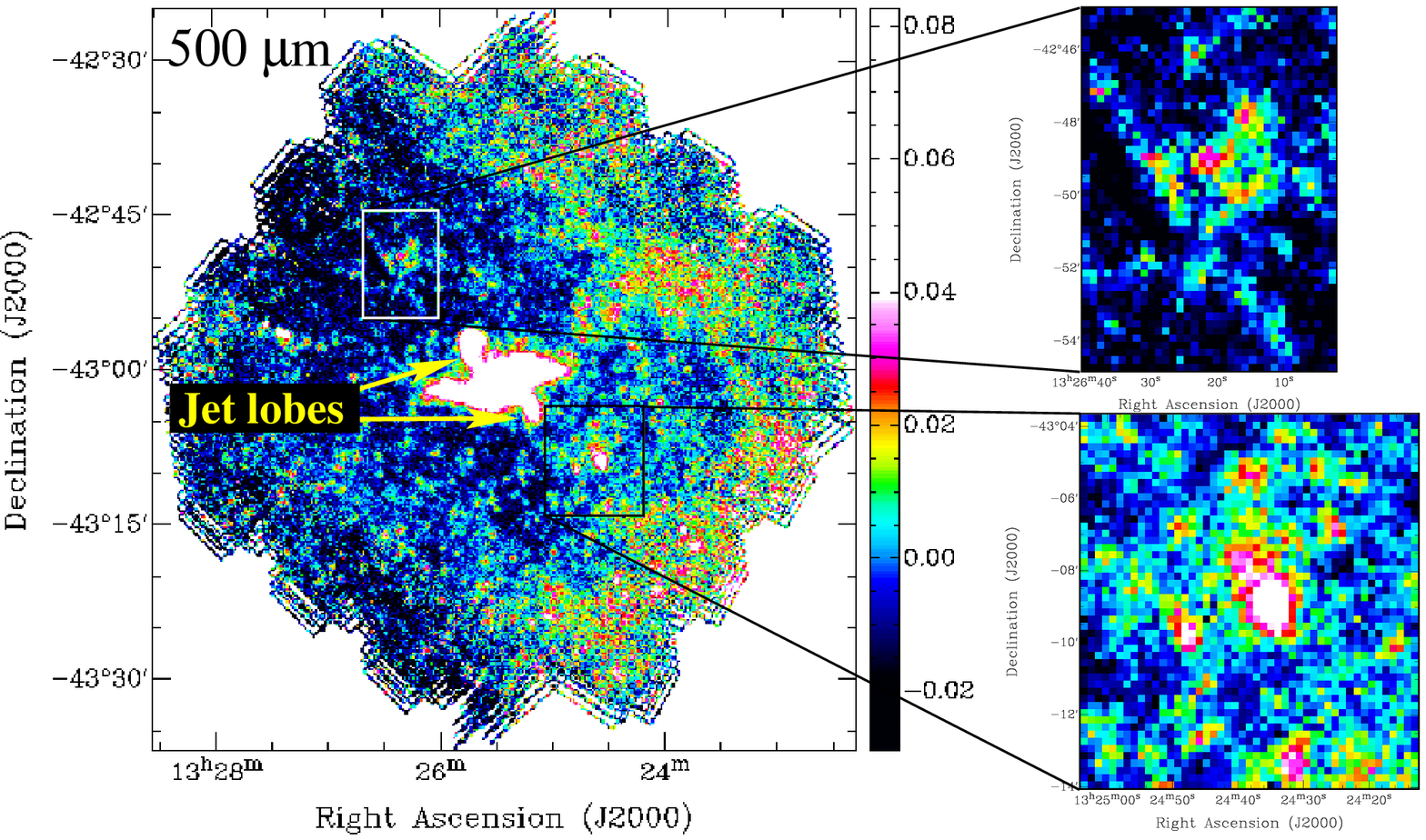}
	\label{fig1a}}
\subfigure{
	\includegraphics[angle=0,width=1.0\textwidth]{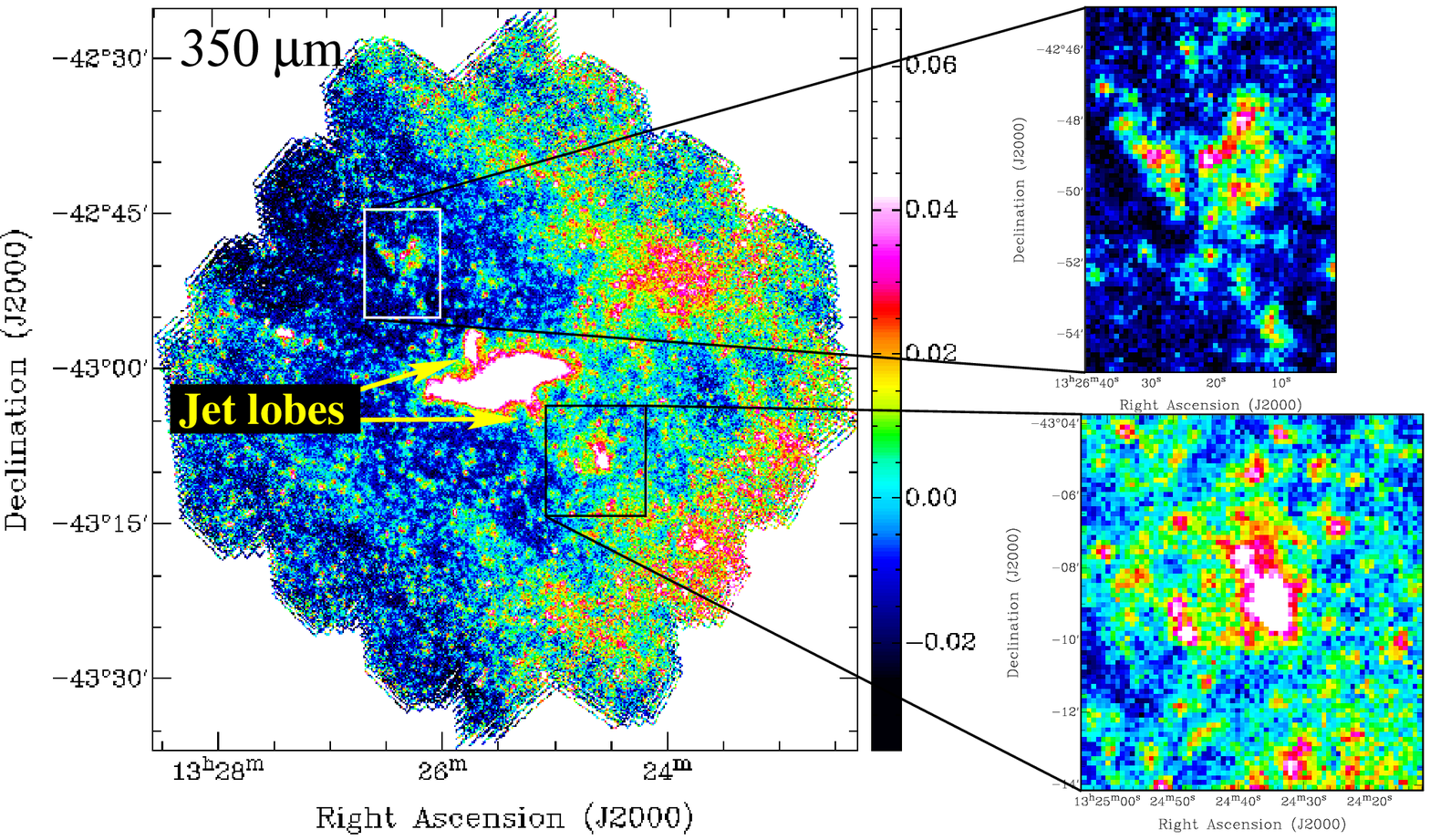}
	\label{fig1b}}
	\caption{{\it Herschel}-SPIRE maps of Centaurus A with insets showing the two dust clouds. {\it top:} 500 \micron, {\it bottom:} 350 \micron, {\it overpage: 250 \micron}. The dusty disk of Cen A dominates the emission at the centre of the image. Synchrotron emission from the inner lobes of the AGN jet are visible as extensions northeast and southwest from the centre of the galaxy. All colour bars are in units of Jy/beam. \label{fig1}}
\end{figure*}

\begin{figure*}
	\includegraphics[angle=0,width=1.0\textwidth]{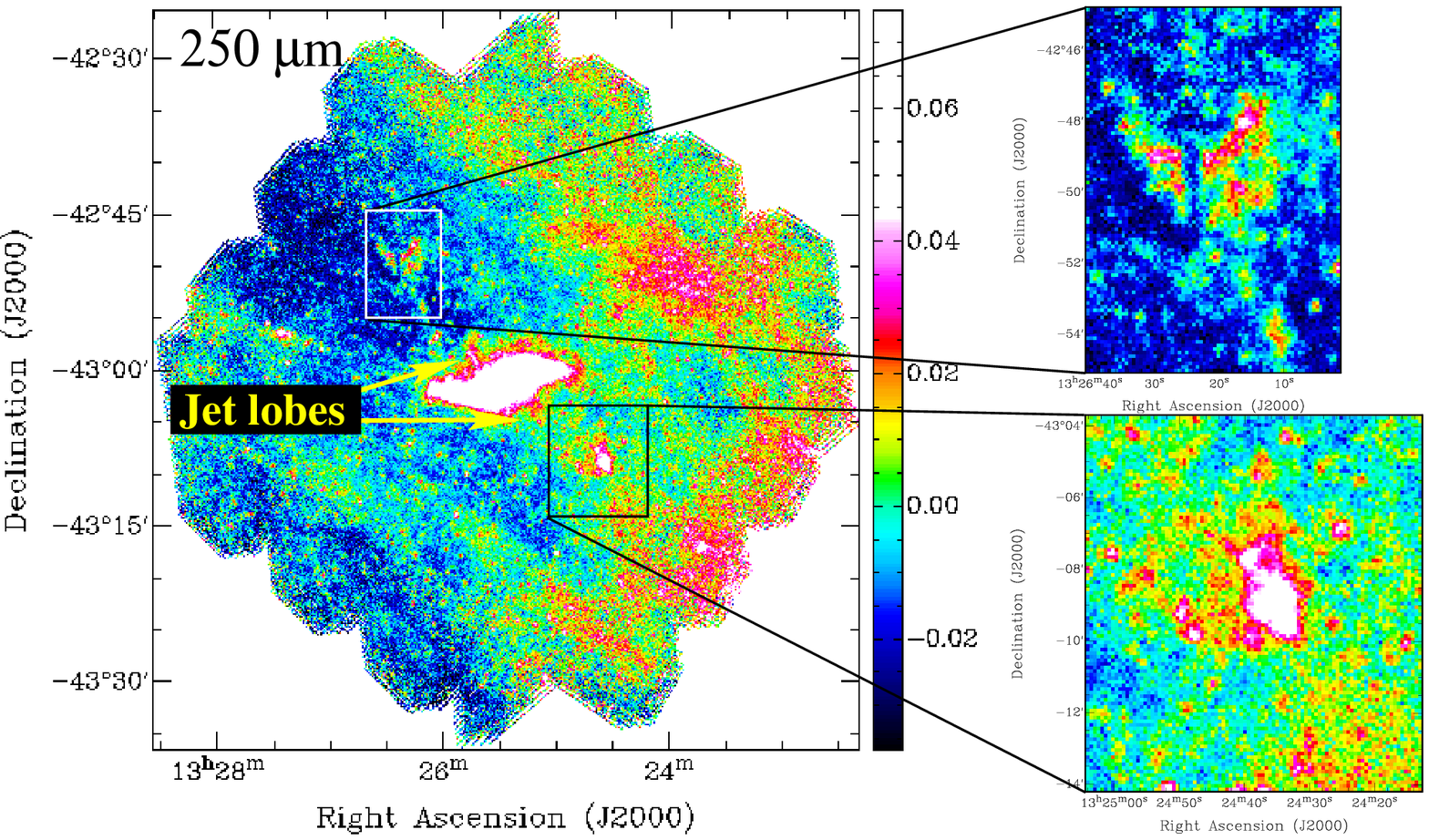}
	\label{fig1c}
	\contcaption{{\it Herschel}-SPIRE maps of Centaurus A: 250 \micron. \label{fig1cont}}
\end{figure*}
	
Fig. \ref{fig1} shows the entire observed field at 250, 350 \& 500 \micron. The prominent dust lane that seems to cut the galaxy in two at optical wavelengths is clearly seen in emission at sub-mm wavelengths in the centre of the field. The inner radio lobes are clearly visible as extensions perpendicular to the dust disk, particularly in the 350 \& 500 \micron\ bands. As stated in the introduction, cold dust peaks longward of 200 \micron. This means that the 350 \& 500 \micron\ bands are situated well on the Rayleigh-Jeans side of the peak of the blackbody curve. Regions of thermal emission are characterised by decreasing flux levels at increasing wavelengths. From comparing the lobes at the different  SPIRE wavelengths, it is clear that the they exhibit the opposite emission trend, suggesting that another emission process must be contributing to these longer wavelengths. Synchotron emission from relativistic electrons spiralling around magnetic field lines in the jet is the most obvious candidate. Synchrotron has a well characterised spectral slope, $\alpha$, such that $S \propto \nu^{-\alpha}$ and thus the longer wavelength {\it Herschel} maps can have significant contributions from synchrotron emission. Synchrotron will be discussed in the section \ref{synch} since its presence will adversely affect the SED fitting routine, which only constrains a thermal spectrum. 

Following the line traced by the jets, we see the two clouds: $\sim$15\arcmin\  NE from the centre of Cen A (13\rah 26\ram 18.876\ras, --42\degree 49\arcmin 32\arcsec) there is a knot of emission about 4\arcmin\  across and $\sim$12\arcmin\  SW from the centre of Cen A, (13\rah 24\ram 34.976\ras, --43\degree 09\arcmin 02\arcsec) there is another cloud $\sim$2\arcmin\  across. Unlike the emission from the inner lobes, the clouds are brighter at shorter wavelengths, suggesting that the sources have a substantial thermal component. The northern cloud is undoubtedly the dust cloud that was detected by Stickel \etal\  (2004), but the southern cloud, to our knowledge, is a new detection.

The morphology of the clouds is very different; the southern cloud is fairly smooth with an elliptical shape, whereas the northern cloud is more flocculent and appears to be two smaller clouds separated by a dust-free lane that runs almost perfectly N--S. 

The Cen A survey field is large enough that the zero level can be affected by the presence of Galactic cirrus, which introduces gradients across the image. Naturally this complicates our estimates of the background in each region. Under the assumption that the emission is optically thin, the southern cloud could be contaminated with more substantial amounts of Galactic cirrus emission than the northern cloud. This explains why the southern source itself appears brighter, but as we shall see in Section \ref{seds} the integrated flux of both clouds is actually fairly similar in each band.

\subsection{{\it Spitzer} images}
\begin{figure*}
\includegraphics[angle=0,height=0.4\textheight]{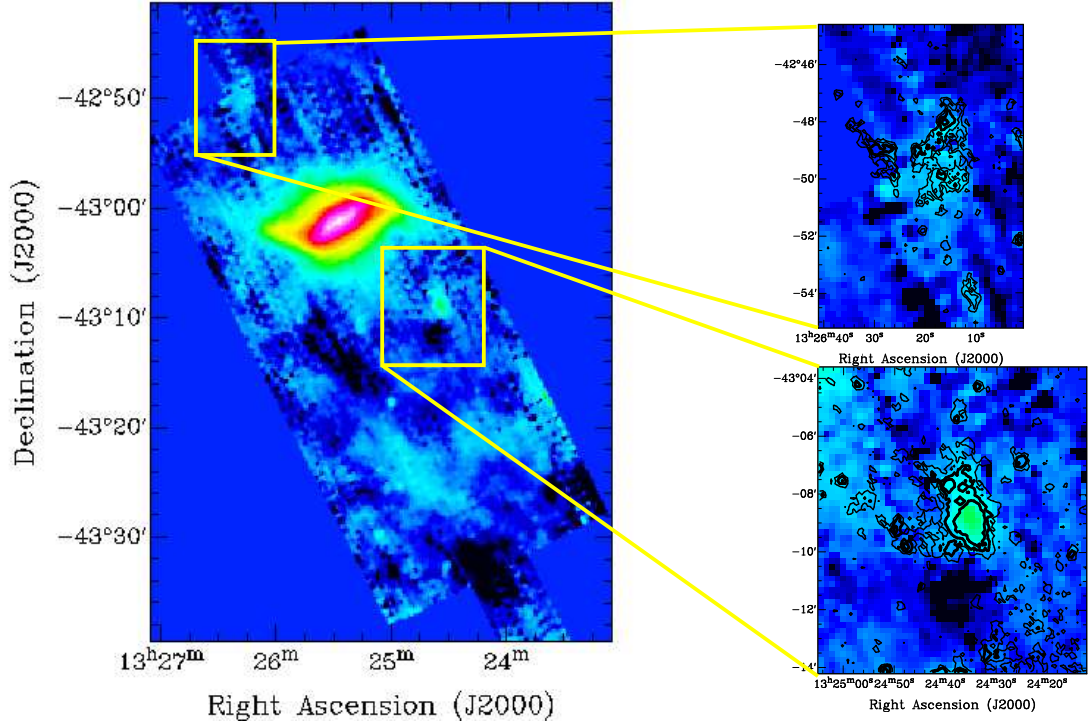}
\caption{{\it Clockwise from left}: {\it Spitzer} 160 \micron\ map of Cen A region. The colourmap has been stretched to enhance diffuse, extended emission. There are a number of visible artefacts, but both dust clouds are visible at the centre of the highlighted regions; the region around the northern cloud with 250 \micron\ contours overlaid at 5, 15, 30 \& 50 mJy/beam; the region around the southern cloud with 250 \micron\ contours overlaid at 15, 30 \& 50 mJy/beam. The 160 \micron\ emission clearly traces the 250 \micron\ emission. \label{fig3}}
\end{figure*}

\begin{figure*}
\includegraphics[angle=0,height=0.4\textheight]{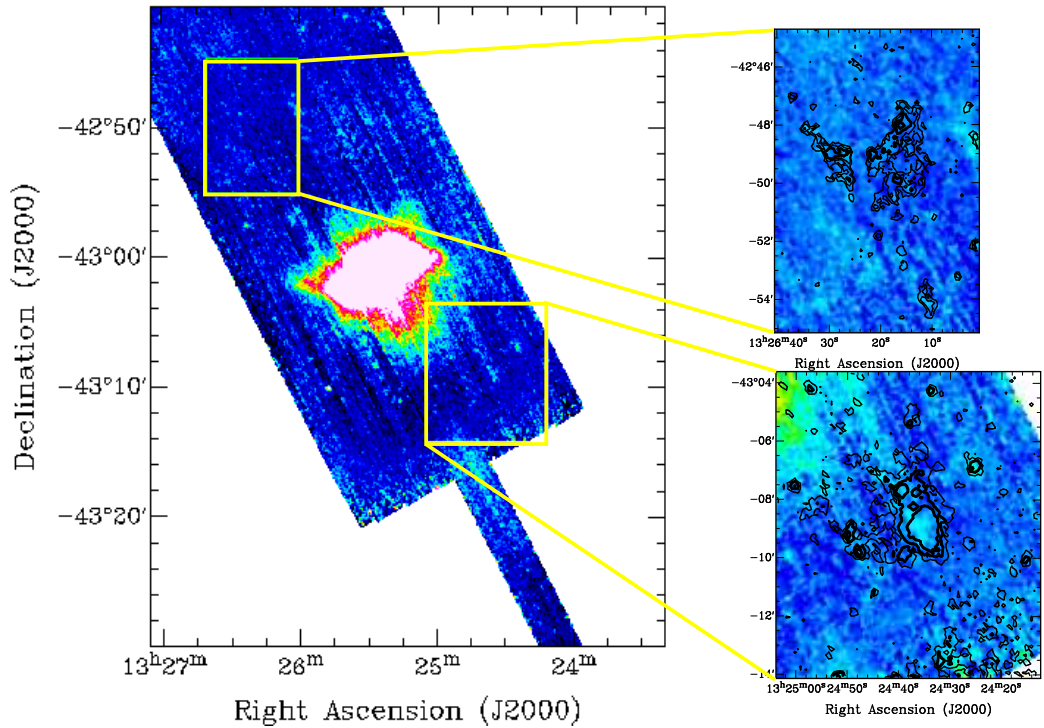}
\caption{{\it Clockwise from left}: {\it Spitzer} 70 \micron\ map of the Cen A region. The colourmap has been stretched to enhance diffuse, extended emission. There are significant stripes present which are artefacts from the observing system that could not be removed during the data reduction process. One of the stripes extends to the position of the southern dust cloud, making flux measurements highly uncertain; the region around the northern cloud with 250 \micron\ contours overlaid at 5, 15, 30 \& 50 mJy/beam; the region around the southern cloud with 250 \micron\ contours overlaid at 15, 30 \& 50 mJy/beam. No 70 \micron\ emission can be unambiguously associated with either cloud. \label{fig4}}
\end{figure*}

\begin{figure*}
	\includegraphics[angle=0,height=0.4\textheight]{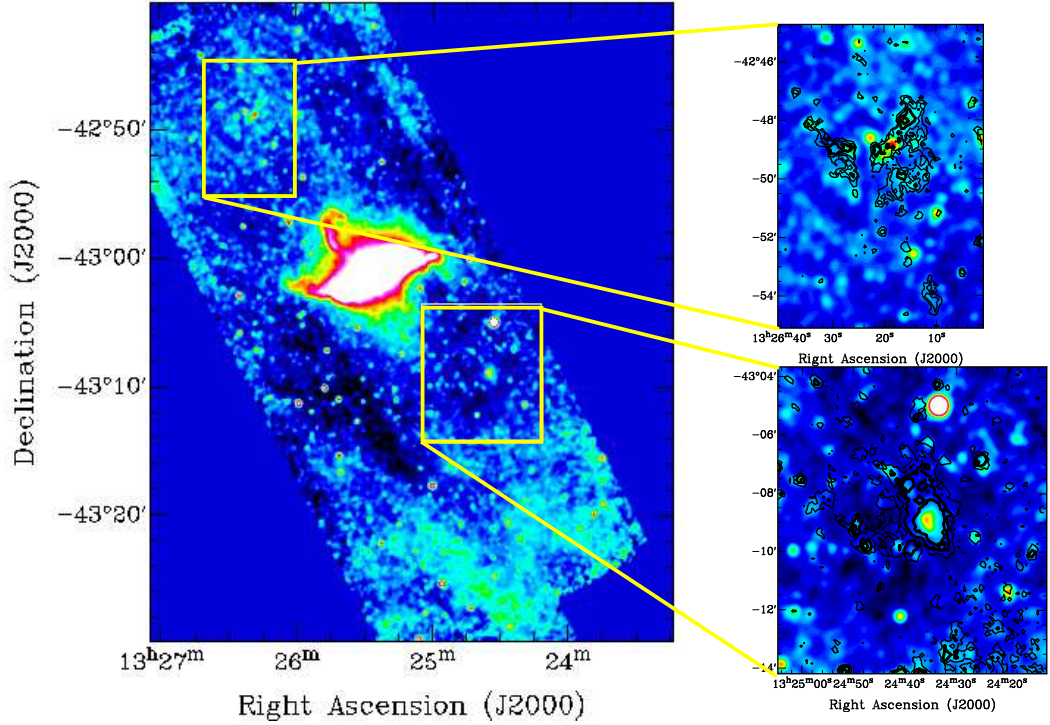}
	\caption{{\it Clockwise from left}:{\it Spitzer} 24 \micron\ map of the Cen A region. The colourmap has been stretched to highlight diffuse extended emission. The northern inner lobe of the jet is just visible as are the dust clouds; sub-image of the northern dust cloud with SPIRE 250 \micron\ contours overlaid at 5, 15, 30 \& 50 mJy/beam; sub-image of the southern dust cloud with 250 \micron\ contours at 15, 30 \& 50 mJy/beam. Both the dust clouds exhibit associated peaks at 24 \micron.\label{fig5}}
\end{figure*}

\begin{figure*}
	\includegraphics[angle=0,height=0.4\textheight]{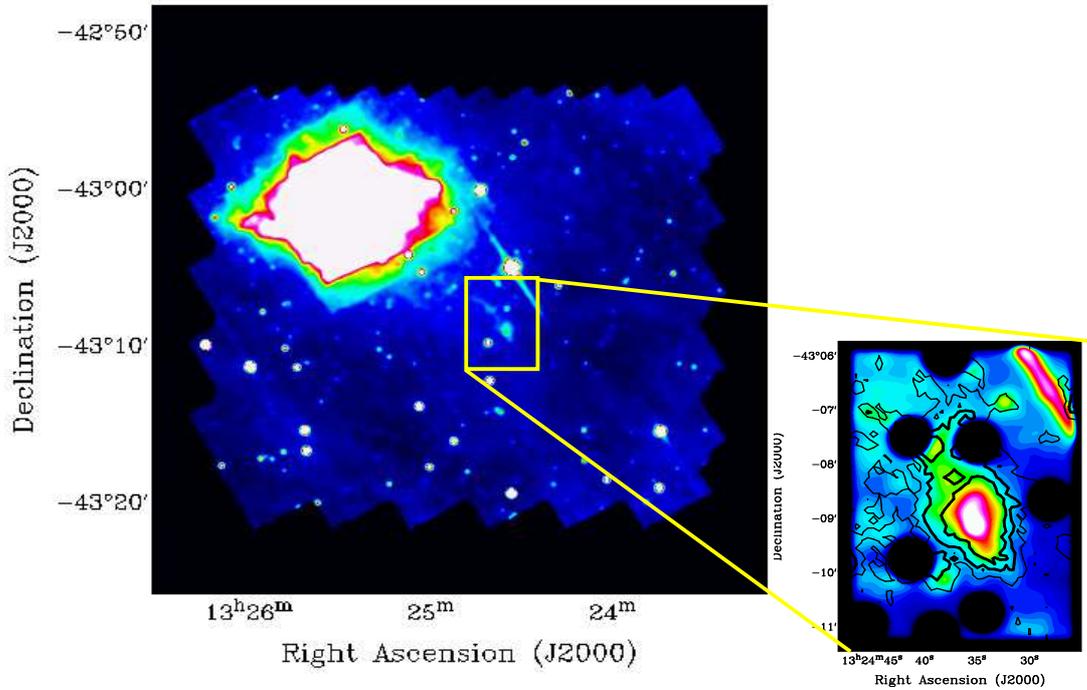}
	\caption{{\it left}: {\it Spitzer} 8 \micron\ map of the region south-west of Cen A. The colourmap has been stretched to enhance diffuse, extended emission. The stellar emission from Cen A dominates the north-eastern corner. The southern dust cloud is visible near the centre of the image; stellar-subtracted 8 \micron\ map of the southern dust cloud with SPIRE 250 \micron\ contours overlaid. The map has been smoothed to the 250 \micron\ beam and the 250 \micron\ contours are: 15, 30 \& 50 mJy/beam. Blank regions indicate foreground stars which have been masked.\label{fig6}}
\end{figure*}

At other wavelengths the clouds show some interesting features. Figs. \ref{fig3}--\ref{fig6} illustrate how the mid-infrared (MIR) and far-infrared (FIR) structures compare with those detected in the sub-mm.   Each image is a 5\arcmin--10\arcmin\ region immediately surrounding each cloud and shows the 250 \micron\ contours overliad on the images at each individual wavelength.
Since 160 \micron\ is close to the peak of the SED for cold dust, one would expect that the morphology exhibited at 250 \micron\ should be simliar to that at 160 \micron\ and this is what we find. In Fig. \ref{fig3} the clouds closely trace the 250 \micron\ emission, although the location of the northern cloud places it close to the edge of the area covered by the MIPS medium scan rate image, hence the area around the northern cloud contains some artefacts related to edge effects at the end of the medium scan rate maps and blank pixels from regions not covered in the fast scan maps. The southern cloud is a strong detection and the extended emission seen at 250 \micron\ is also seen at 160 \micron.

\begin{figure*}
\includegraphics[angle=0,height=0.4\textheight]{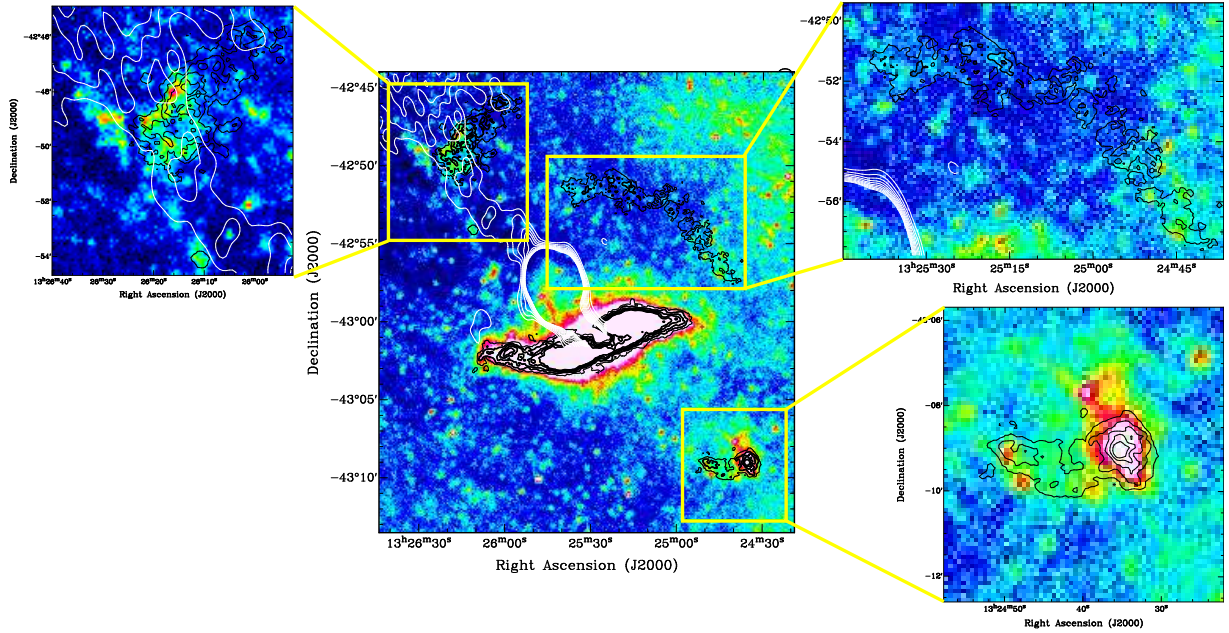}
	\caption{Multiwavelength observations of the immediate surroundings of Cen A. SPIRE 250 \micron\ false colour image with 20cm continuum emisssion (white contours) and \hi\ column density (black contours). The image colour scale has been stretched to enhance low surface brightness features, contour levels are as follows: 20cm -- 12.5$\times$ 1, 2, 3, 4, 5, 6, 7, 8 \& 9 mJy/beam; \hi --  1, 5, 10, 15 \& 20 $\times 10^{20}\ \rm{cm}^{-2}$. The image is centred on Cen A and highlights the regions of the \hi\ ring.\label{fig7}}
\end{figure*}
\begin{figure*}
\includegraphics[angle=0,height=0.35\textheight]{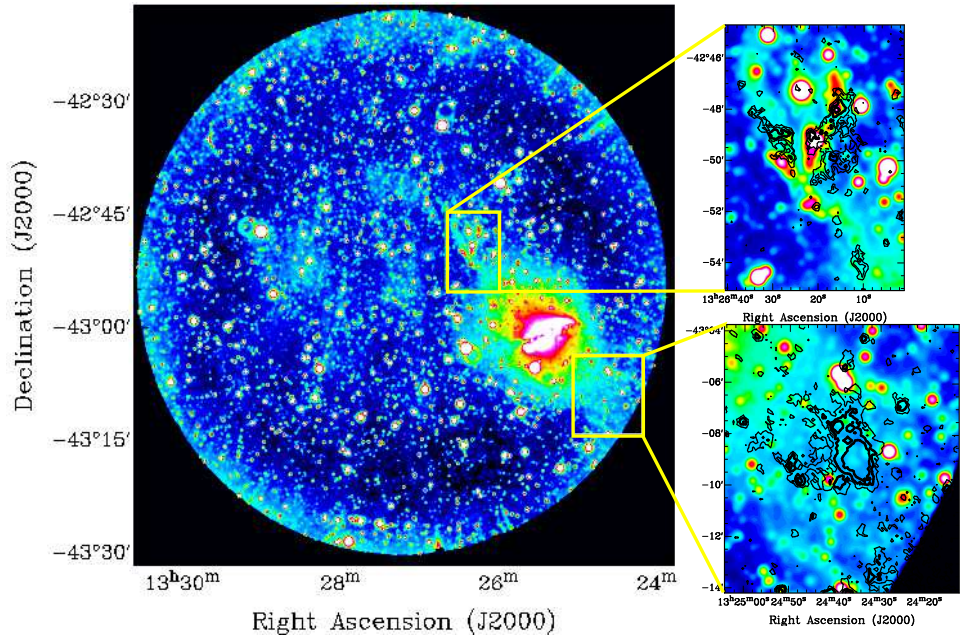}
\caption{{\it Clockwise from left}: GALEX NUV map of the Cen A region. The map has been smoothed to the SPIRE 250 \micron\  beam size and the colour scale has been stretched to enhance low-surface brightness, extended features. There is a lot of extended emission corresponding to the stellar population of Cen A. A jet-like feature can be seen extending far beyond the stellar radius to the north of Cen A; sub-image of the northern dust cloud with SPIRE 250 \micron\ contours overlaid at 5, 15, 30 \& 50 mJy/beam. The region exhibits a large amount of UV coincident with the dust emission; sub-image of the southern dust cloud with 250 \micron\ contours at 15, 30 \& 50 mJy/beam. This dust cloud has no detectable UV emission associated with it. \label{fig8}}
\end{figure*}
\begin{figure*}
\includegraphics[angle=0,height=0.35\textheight]{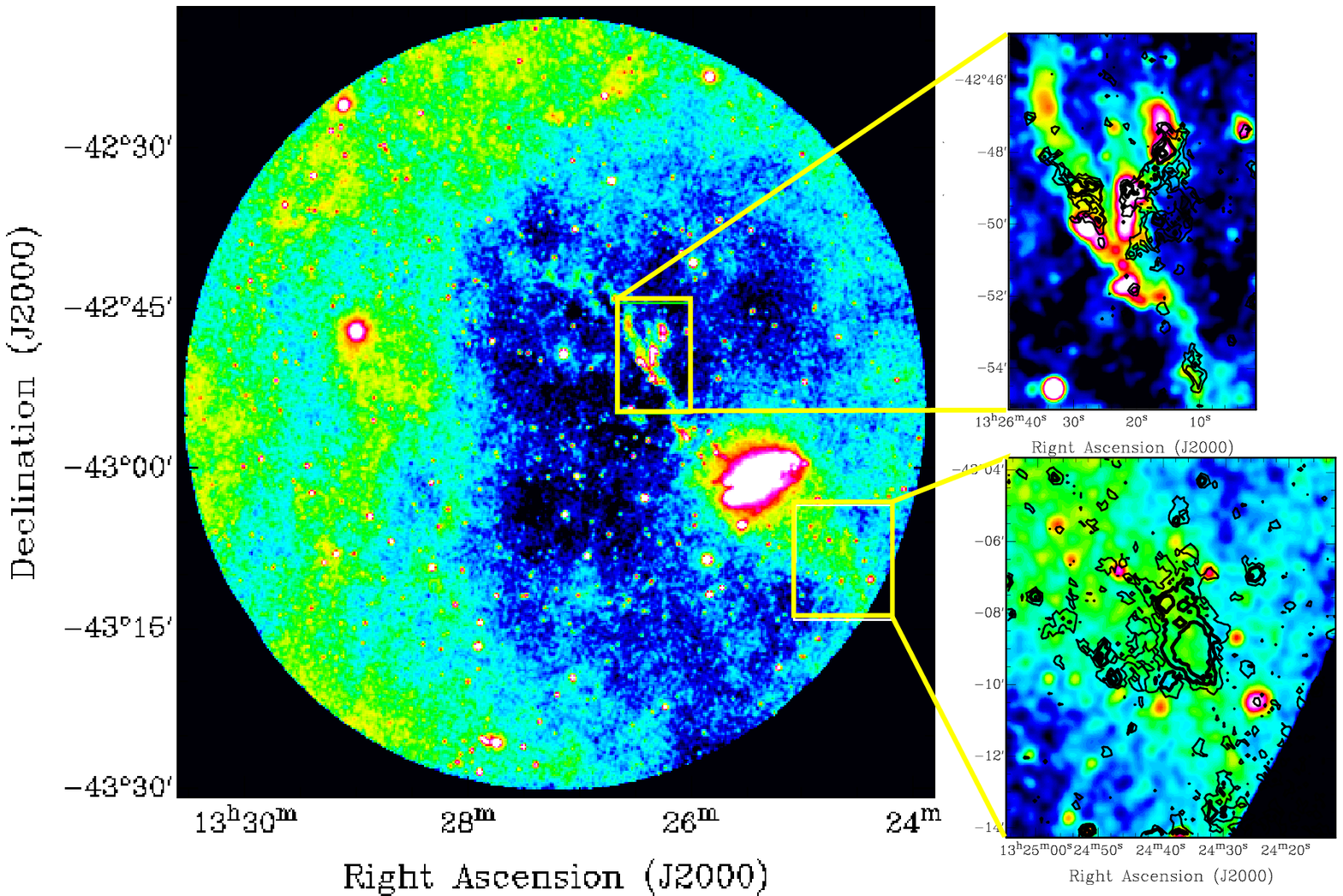}
\caption{{\it Clockwise from left}: GALEX FUV map of the Cen A region. The map has been smoothed to the SPIRE 250 \micron\  beam size and the colour scale has been stretched to enhance low-surface brightness, extended features. As well as the extended emission from Cen A, Galactic cirrus can be seen in the eastern half of the map. The jet-like feature noticed in the NUV is also visible here; sub-image of the northern dust cloud with SPIRE 250 \micron\ contours overlaid at 5, 15, 30 \& 50 mJy/beam. The region exhibits a large amount of UV coincident with the dust emission; sub-image of the southern dust cloud with 250 \micron\ contours at 15, 30 \& 50 mJy/beam. This dust cloud has no detectable UV emission associated with it. \label{fig9}}
\end{figure*}

If the dust is cold, the spectrum drops off rapidly at wavelengths shorter than $\sim$160 \micron. The expected 70 \micron\ emission is a few orders of magnitude lower than that at 160 \micron\ so a lack of any detection at 70 \micron\ would indicate a cold object. While it seems there is a peak in Fig. \ref{fig4} at the location of the southern cloud, it happens to be on top of a streak in the image which is a data reduction artefact. Because of this, we consider the southern detection spurious and have only used data from the 70 \micron\  maps to assess upper limits. Using the rms in the vicinity of the clouds leads to an upper limit of $\sim$200 mJy.

One of the most interesting results to come out of the {\it Spitzer} Nearby Galaxy Survey was the tight correlation between the 24 \micron\ emission and the atomic hydrogen spectral line emission from Paschen-$\alpha$, P$\alpha$ (Calzetti \etal, 2005, 2007; Prescott \etal, 2007; Zhu \etal, 2008, Kennicutt \etal, 2009). This suggests that the 24 \micron\ band is a particularly good tracer of star forming regions (although see Bendo \etal, 2010 for alternative sources of 24 \micron\ emission). In Fig. \ref{fig5} we see that both of the clouds show 24 \micron\ emission coincident with the peaks in the 250 \micron\ emission, suggesting a common source for both bands.

Emission from galaxies in the 8 \micron\ band of Spitzer is dominated by broad line emission from polycyclic aromatic hydrocarbons (PAHs). There is also a correlation between the 8 \micron\ PAH emission (corrected for stellar emission) and P$\alpha$, however the correlation is not as tight as for 24 \micron. The relationhip is non-linear and more dependent on metallicity than the 24 \micron\ band, so it is not as reliable to associate 8 \micron\ emission with star formation (e.g. Calzetti \etal, 2005; Bendo \etal, 2008). Unfortunately the northern cloud falls outside the survey field in the 8 \micron\ band, but the southern field shows some point-like sources embedded in a diffuse, extended cloud. When smoothed to the 250 \micron\ beam, the point sources blend together so that the emission appears as a cloud which closely traces the 250 \micron\  contours, again suggesting a common origin.

\subsection{Radio images}

Fig. \ref{fig7} shows the location of the northern jet (white contours) and the \hi\ ring (black contours) overlaid on the 250 \micron\ false colour image.  
The systemic velocity of the gas is beyond the extent of Galactic \hi, which indicates that the gas is part of the Cen A system. The dust emission correlates well with the \hi\ contours, suggesting that the dust is associated with the \hi\ in Cen A as opposed to being a coincidental projection of Galactic cirrus (c.f. M81: Davies \etal, 2010).

Previous comparisons of the \hi\ velocity dispersion of the southern tip of the northern cloud match the velocity range of ionised gas just to the east of the cloud (Oosterloo \& Morganti, 2005). This kinematic agreement combined with the spatial coincidence is a good indicator that an interaction is taking place between the jet and the \hi\ cloud. The dust is also coincident with the jet, at least in projection. If the dust is well mixed with the gas, as we'd expect, it would imply that the dust is also involved in the jet-cloud interaction. It is also interesting to note that the western part of the \hi\ ring shows no corresponding enhancement in the sub-mm. Possible explanations for this are discussed in Section \ref{merger}.

\subsection{Ultraviolet images}
Both of the GALEX images (Figs. \ref{fig8} \& \ref{fig9}) have been smoothed to the resolution of the SPIRE 250 \micron\ beam to help bring out some of the more diffuse UV structure. UV emission from the stellar population can be clearly seen in both images. A large amount of emission is present in the eastern half of Fig. \ref{fig9} which is mostly likely scattered light from Galactic cirrus. Also striking is the jet-like feature extending from Cen A to the north, far beyond the optical extent of the galaxy. Focusing on the clouds themselves reveals a stark difference between them. The northern cloud clearly has UV emission coincident with it, but the emission is mostly restricted to the edges of the cloud. It may be a chance alignment, but is also what one would expect if the UV emission is embedded within, and obscured by, the dust cloud. The southern cloud, however, shows no sign of UV emission in either of the GALEX bands. It is possible that there is no UV originating from the southern dust cloud which would indicate the lack of any star formation in the vicinity. Since UV emission is highly attenuated by dust grains, this could also indicate that UV is present but has been obscured by high column density gas and dust. In the next section, we combine the {\it SPIRE}, {\it Spitzer} and radio data to assess whether the dust cloud emission is purely thermal or contaminated by synchrotron.

\subsection{Synchrotron Contamination}
\label{synch}
Since the inner lobe is particularly bright in the 500 and 350 \micron\  image, we must consider whether or not the fluxes are contaminated by synchrotron radiation from the jet. As mentioned in Section \ref{spireimages} synchrotron radiation has a power-law emission spectrum. It is therefore possible to estimate the amount of contamination at the SPIRE wavelengths by measuring the clouds' fluxes from lower frequencies (e.g. from radio maps) and extrapolating to the SPIRE wavelengths.
 
Ideally the spectral index would be measured from a combination of radio maps, but the map of Morganti \etal\  (1999) is the only radio map with sufficient spatial resolution that extends to the distance of the clouds. In this case we are reduced to assuming a value for the spectral slope. The most extreme case would be to assume that the spectral index is as flat as possible, i.e. $\alpha = 0.5$ (Begelman, Blanford \& Rees, 1984). We measure a radio flux inside the aperture used for the SED fitting of 270$\pm$60 mJy at 1.4 GHz. Even with the flat spectral index this only produces a flux of a few mJy at the 500 \micron\ band. Hence, there is no significant contamination by the jet in the SED of the northern cloud.

To our knowledge no radio maps of sufficient resolution and sensitivity exist for the region covering the southern cloud. Since there is no equivalent to the Northern Middle lobe connecting the southern inner lobe to the giant outer lobe, the amount of synchrotron must be even lower than in the northern cloud. Having rejected synchrotron as a possible contaminant we model the fluxes under the assumption that the emission is thermal in origin.

\subsection{SED fitting}
\label{seds}
Stickel \etal\  (2004) fit the FIR SED for the northern cloud using a modified blackbody spectrum, but their analysis was limited to two points on the Wein side of the spectrum (150 \& 200 \micron) plus another upper limit (90 \micron). With so few data points the peak was poorly constrained. With the combination of MIPS and SPIRE data, we now have 5 points straddling the peak, which should enable us to fit a modified blackbody in the same manner, but with a higher degree of assurance.

We adopted the SED fitting routine as used in Smith \etal\  (2010), in which the uncertainties in the fitted dust mass and temperature are derived from several thousand Monte Carlo simulations. The benefit of the simulations is the determination of the error distributions for the mass and temperature which are not necessarily Gaussian, thus lending a higher degree of robustness to the quoted uncertainties. The fitting routine allows the user to specify the emissivity index, $\beta$ as a free parameter, but we found that the simulations did not converge, so $\beta$ was fixed at 2. The dust mass-opacity coefficient used was 0.192 m$^2$ kg$^{-1}$ at 350 micron (Draine \& Li, 2001).

Before fitting the SED, the images were smoothed to the resolution of the SPIRE 500 \micron\ beam. An aperture was defined for each cloud based on the highest signal-noise map (250 \micron) and these apertures were used across all bands. The lack of coverage in the 160 \micron\ image in the north limited our ability to measure flux over the whole cloud, but only a small section of the eastern-most part of the cloud was lost and it is unlikely to make a significant difference to the final result. 

The measured fluxes are shown in Table \ref{tbl1}. Quoted uncertainties are larger than the nominal calibration uncertainties for each instrument, but given the amount of foreground contamination against which the fluxes were measured, we feel these larger error bars are justified. It was apparent that even after smoothing, the quality of the 70 \micron\ images was only good enough to provide upper limits, but this still gave stricter constraints than omitting the 70 \micron\ data altogether.

\begin{table}
  \centering
  \caption{FIR-sub-mm flux measurements for the two dust clouds. Uncertainties are shown in brackets and are 20\% for SPIRE fluxes and 30\% for Spitzer measurements.\label{tbl1}}
  \begin{tabular}{lccccc}
      \hline\\
    Cloud & $S_{500}$ & $S_{350}$   & $S_{250}$  & $S_{160}$  & $S_{70}$  \\
      & Jy & Jy & Jy & Jy & Jy \\
      \hline\\
      North & 0.33 (0.07) & 0.8 (0.2) & 0.8 (0.2) & 0.7 (0.1) & $<$ 0.2\\
	  South & 0.37 (0.07) & 0.9 (0.2) & 1.0 (0.2) & 1.5 (0.5) & $<$ 0.2\\
      \hline\\
  \end{tabular}
\end{table}

\begin{figure}
	\includegraphics[angle=0,width=0.5\textwidth]{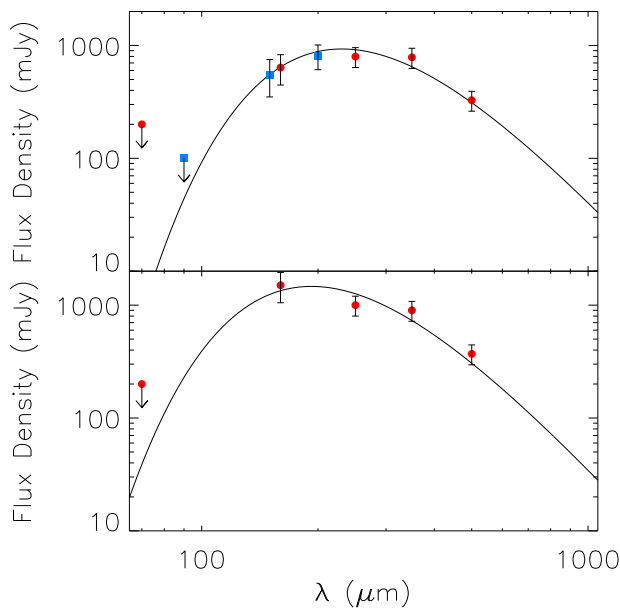}
	\caption{Modified blackbody, single-temperature fits to the FIR--sub-mm data assuming a $\beta=2$ emissivity. Data points from this analysis are shown as red dots and the data points of Stickel \etal\  are shown as blue squares. Arrows indicate upper limits. {\it Top:} northern cloud. {\it Bottom:} southern cloud. \label{fig10}}
\end{figure}

The SEDs for the dust clouds are presented in Fig. \ref{fig10}. The SED for the northern cloud compares well with the SED of Stickel \etal\  (2004), and we derive a similar dust temperature: T = $12.6^{+1.1}_{-1.2}$ K. We recover more mass than Stickel \etal, $\rm{log} (M_{dust} / M_{\odot}) = 5.8^{+0.2}_{-0.2}$, most likely because of the increased wavelength coverage provided by {\it Herschel}. For the southern cloud we measure a dust mass and temperature of $\rm{log} (M_{dust} / M_{\odot}) = 5.6^{+0.2}_{-0.2}$, T = $15.1^{+1.7}_{-1.6}$ K. Combining the dust masses with the gas masses measured by Charmandaris \etal\  (2000), we derive gas-dust ratios of 121 \& 110 which are typical of late-type, star-forming galaxies. Having established a thermal origin for the sub-mm emission in these clouds, we now discuss their origin and heating mechanisms.

\section{Discussion}
\label{discussion}
\subsection{Origin of the clouds}
\subsubsection{Galaxy merger}
\label{merger}

A comparison of the distribution of the atomic gas and the dust (Fig. \ref{fig7}) reveals that the only detected dust emission, that can be unambiguously associated with the \hi\ ring, is at the position of the two clouds. The lack of sub-mm emission in the rest of the gas ring has three possible explanations: either there is no dust in other parts of the ring, the column density of dust falls below the SPIRE detection limit, or the dust is present but too cold to emit at the SPIRE wavelengths.

If the galaxy that merged with Cen A was a late-type, it would be expected that the dust is well mixed with the gas in the \hi\ ring. It is unlikely therefore, that there is no dust anywhere else in the orbitting gas. This then leaves the question of whether the dust is too diffuse or too cold.

Assuming that the dust in the HI cloud to the west is at the same temperature as
 the two detected clouds, the SPIRE detection limit can be used to estimate the limiting column density of dust within it. Apertures were used to define the local rms in the vicinity of the cloud and this was adopted as the noise level. The source region was defined from the \hi\ map of Struve (2010), using an \hi\ column density of $10^{19}$ cm$^{-2}$. The 500 \micron\ map was the most sensitive in the region of the western cloud. Despite having a slightly larger instrumental noise value, this wavelength is less sensitive to Galactic cirrus which dominates the background in this region. The rms pixel value in the vicinity of the \hi\ cloud was 10.1 mJy/beam. This gave a 3-$\sigma$ flux upper limit of 66.6 mJy over the extent of the cloud. For a 14K cloud with $\beta =2$, this resulted in an upper limit for the dust mass of $M_{dust} < 9.4 \times 10^5 M_{\odot}$, or a corresponding gas-dust ratio of $M_{gas}/M_{dust} > 63$. This is not a particularly strong lower limit for the gas-dust ratio, and suggests that the dust is simply too diffuse to be detected by SPIRE. Without a strong upper limit on the gas-dust ratio there is little use in exploring the impact of a significant amount of colder dust in the western \hi\ cloud.
 
These findings are consistent with having dust well mixed throughout the gas ring and the gas-dust ratio is typical of late-type galaxies as is the ratio of the atomic to molecular gas. This supports the scenario that the dust and gas originated in the galaxy that merged with Cen A to form the dusty disk at the centre of Cen A.

The northern cloud is coincident with the high-velocity \hi\ gas which has been linked to the jet-induced star forming region nearby. Opportunities to explore jet-induced star formation in the kind of detail that Cen A offers are rare, so it is interesting to explore the possibility that this local star formation is responsible for the existence of the northern dust cloud.

\subsubsection{Jet-induced Star Formation}
If the dust originated from the stars in the vicinity of the clouds, we can use a simple argument to determine whether or not the SFRs are sufficiently high to have formed the mass of dust we observe. Stardust is formed in the cool, stellar winds of low-intermediate mass stars (LIMS) on timescales of $0.5-1\,\rm Gyr$ and in supernovae (SNe) on timescales of tens of Myr. In the Milky Way, the major dust source is presumed to be the former, with LIMS injecting dust at a rate of $\sim$2 $\times 10^{-3}\,\rm M_{\odot}\,yr^{-1}$ (e.g. Whittet 2003). Recent FIR observations of Galactic and extragalactic SN remnants with {\it Spitzer} and {\it Herschel} (e.g. Rho \etal\ 2008; Otsuka \etal\ 2010; Barlow \etal\  2010) suggest SNe contribute a dust mass-loss rate of $< \rm 1\times 10^{-3} \,M_{\odot}\, yr^{-1}$ (corresponding to $<0.1\,\rm M_{\odot}$ of dust formed per SNe).  However, as SN shocks are also responsible for destroying dust in the ISM via sputtering, it is not yet clear how much of the dust observed in remnants survive their journey into the ISM.

We can estimate the stellar dust injection rate needed to produce the observed dust mass in the Cen A clouds by comparing the rate at which dust grains are destroyed in the ISM.  The timescale for dust destruction can be estimated using (McKee 1989):

\begin{equation}
t_{\rm
  des} = {M_d \over{R_{\rm SN} M_{d,s}}} = {M_d \over{{\rm 0.05} SFR}}
\label{eq:dest}
\end{equation}

where $M_d$ is the dust mass, $R_{\rm SN}$ is the SN rate, $M_{d,s}$ is the mass of dust shocked and destroyed by one SN blast wave ($\sim$3 M$_{\odot}$ for shocks with speeds $v>100\,\rm km\,s^{-1}$ - Dwek \& Cherchneff 2011).  Using Eq.~\ref{eq:dest}, dust is destroyed on timescales of $\rm \tau_{des} \sim$400 $\rm \,Myr$ in the Milky Way (Jones \etal\  1994; Tielens 1998). The estimate requires knowledge of the SFR, and a number of prescriptions exist to estimate the SFR for the northern cloud based on its luminosity at 24 \micron, $NUV$ and $FUV$. For completeness the SFRs for the southern cloud are also calculated, but the reader is reminded that there is no evidence of a connection between the southern jet and this dust cloud so in this case the assumed star formation is not jet-induced.

Using the prescription in Rela\~no \etal\  (2007) as presented by Calzetti \etal\ (2010) the northern cloud exhibits $S_{24} = 9 $ mJy, which corresponds to a $SFR_{24} = 1.4 \times 10^{-3} {\rm M}_{\odot} {\rm yr}^{-1}$. For the southern cloud we measure $S_{24} = 7$ mJy. This corresponds to a $SFR_{24} = 1.2 \times 10^{-3} {\rm M}_{\odot} \rm{yr}^{-1}$.

Iglesias-P\'aramo \etal\  (2006) provide recipes for calculating SFRs in the UV bands of GALEX. The UV luminosities must first be corrected for Galactic extinction. Galactic extinction magnitudes, $E(B-V)$, were provided by the GALEX, MAST website using objects identified close to ($<30$\arcsec) the cloud positions within the Cen A field. For the northern cloud: $E(B-V) = 0.102$ and for the southern $E(B-V) = 0.138$. The conversion from $A_B$ to $A_{UV}$ was calculated using the method described in Wyder \etal\  (2007). The attenuation  values are: $A_{UV} = 0.840$ for the northern cloud and $A_{UV} = 1.055$ for the southern cloud. The measured NUV and FUV magnitudes were then corrected by these values. 

For the southern cloud, the 5-$\sigma$ limits based on the error in the background estimates, corrected for Galactic extinction, were used to provide upper limits. The extinction-corrected luminosities were then used to calculate the SFRs. For the northern cloud we derive a SFR$_{FUV} = 3.8 \times 10^{-3} {\rm M}_{\odot} \rm{yr}^{-1}$ and SFR$_{NUV} = 2.4 \times 10^{-3} {\rm M}_{\odot} \rm{yr}^{-1}$ for the southern cloud SFR$_{FUV} < 2 \times 10^{-4} {\rm M}_{\odot} \rm{yr}^{-1}$ and SFR$_{NUV} < 5 \times 10^{-5} {\rm M}_{\odot} \rm{yr}^{-1}$. All the luminosity measurements and associated SFRs are shown in Table \ref{tbl2}. For reference, the SFR derived from Mould \etal\  (2000) is also shown.

The SFR values for the northern cloud as judged from four independent tracers differ by factors of a few. The SFRs broadly agree, given the different caveats and scatter that are associated with each SFR estimator. The low SFR for the northern cloud implies dust in this environment could survive for $2\,\rm Gyr$.  The dust injection rate from stars for this cloud would therefore need to be $\rm M_d/\tau_{des} \sim 10^{-4} \,M_{\odot}\,yr^{-1}$ to produce the total dust mass.  From the low SFR ($\sim 10^{-3}\,\rm M_{\odot}\,yr^{-1}$), we would expect a dust injection rate from LIMS and SNe (scaled from the Milky Way parameters and SFR) of $10^{-5} \rm \,M_{\odot}\,yr^{-1}$; this is an order of magnitude lower than required. At this rate, it would take more than 30\, Gyrs to build up the $4 \times 10^5\,\rm M_{\odot}$ of dust observed by {\it Herschel} with a stellar mass-loss source of dust.

The dust clouds detected here cannot be explained by replenishment of the dust via stellar mass loss; the destruction timescales would need to be more than an order of magnitude longer than currently estimated and the induced star formation would need to be replenishing dust for an unfeasibly long time. This reinforces the notion that the dust most probably originated in the galaxy that merged with Cen A and was re-distributed along with the gas as the galaxy was torn apart. 

Simulations of merger events have demonstrated that the gas and dust are preferentially stripped, so the lack of a stellar enhancement in the \hi\ ring is not unusal. It does, however, raise the question of what possible heating source could be responsible for heating the clouds to produce their sub-mm emission.

\begin{table}
  \centering
  \begin{minipage}[c]{83mm}
  \caption{IR and UV star formation rates for the dust clouds.\label{tbl2}}
  \begin{tabular}{lcc}
      \hline\\
    Cloud &          North &     South \\          
      \hline\\
S$_{24}$ (Jy) & 0.009 & 0.007\\
L$_{24}$ ($\times10^{39}$erg s$^{-1})$ & 1.69 & 1.36\\
SFR$_{24} ($M$_{\odot}\rm{yr}^{-1}$) & 0.00143 & 0.0012\\
m$_{NUV}$ & 15.3 & $>20.96$\\
m$_{FUV}$ & 16.1 & $>19.06$\\
$E(B-V)$ & 0.102 & 0.128\\
L$_{NUV}$ ($\times10^{40}$erg s$^{-1})$ & 5.92 & $<0.06$\\
L$_{FUV}$ ($\times10^{40}$erg s$^{-1})$ & 4.21 & $<0.15$\\
SFR$_{NUV}$ (M$_{\odot}\rm{yr}^{-1}$) & 0.0024 & $<5\times10^{-5}$\\
SFR$_{FUV}$ (M$_{\odot}\rm{yr}^{-1}$) & 0.0038 & $<2\times10^{-4}$\\
 SFR$_{M00}$\footnote{Mould \etal (2000) based on a the maximum age of detected OB associations and total number of detected OB associations assuming a Salpeter mass function} (M$_{\odot} \rm{yr}^{-1}$)&      0.0012&     ---\\
   \\
      \hline\\
  \end{tabular}
  \end{minipage}
\end{table}

\subsection{Dust heating mechanisms}

The dust clouds reside in a complex environment. They exist on the very outskirts of the stellar extent of Cen A and heating by the stellar population will have some effect. The region around Cen A is also rich in X-ray emission and the northern cloud appears to lie in the projected path of the AGN jet and also close to the beam of ionising radiation discovered by Morganti \etal\ (1991). We now consider each of these heating mechanisms and discuss their relative contribution. 

\subsubsection{Heating by aging stars}
\label{temi}
The model of Temi \etal\ (2003) predicts temperatures of dust grains of different sizes as a function of radius for early type galaxies such as Cen A. The model  environment consisted of X-rays from a hot ($>10^7$ K) ISM and starlight from the evolved stellar population. From Fig. 2 of Temi \etal\ (2003) the expected dust temperature at the distance of the dust clouds is 14-16 K for grains of size $\sim 1$\micron, which is entirely consistent with our measurements and also the previous measurement of Stickel \etal\ (2004). For smaller grains, the temperature profiles are too high to account for the low temperatures we see, so the clouds must be dominated by large grains if the Temi model is representative of the grain population. The temperature profiles from grains of this size are dominated by heating from the aging stars, with electron-heating becoming more important for smaller grains. From this we infer that the clouds are most likely heated by the old stellar population. 

It is curious that in such a complicated environment, a single heating mechanism can account for the grain heating. We now examine the alternative possible mechanisms and attempt to explain why they are not credible.

\subsubsection{Beamed radiation}
Ionisation of the material close to the dust clouds in Cen A by a beamed nuclear source was first proposed by Morganti \etal\  (1991, 1992). Through extensive modelling incorporating multi-cloud photoionisation, they were able to reject  other scenarios such as heating from a diffuse X-ray halo or cosmic ray heating in favour of the beamed, ionising photons, emanating from a source embedded deep within the obscured nucleus of Cen A.
We have used their estimate of ionising photon flux to calculate the temperature of a dust grain at the distance of the clouds. 

In the absence of attenuation, the total energy incident on a grain is simply given by:
\begin{equation}
	Q_{heat} = q h\nu_{H\alpha} \Omega_{grain} \label{eqn1}
\end{equation} 
where $q$ is the ionising photon flux (photons s$^{-1}$ sr$^{-1}$), $h\nu_{H\alpha}$ is taken as an estimate of the minimum energy of the ionising photons and $\Omega_{grain}$ is the solid angle of a single grain ($\sim$3.3 $\times 10^{-55}$ sr) at a distance of the dust clouds ($\sim$15 kpc).

This incident power is then absorbed completely by the grain, and radiated away:

\begin{equation}
	Q_{cool} = 4 \pi a_{X}^2 \sigma_{SB} T_d^{4} \langle Q_{abs} \rangle \label{eqn2}
\end{equation} 

where $a_{X}$ is the cross-sectional area of a grain (a spherical grain diameter of 1 \micron\ is assumed) and $Q_{abs}$ is the absorption co-efficient and we use the approximation $Q_{abs}\approx 1.35 \times 10^{-5} T_d^2 a_X$ following Temi \etal\  (2003) and Draine \& Lee (1984). 

By conservation of energy, Eqns. (\ref{eqn1}) \& (\ref{eqn2}) must balance. So if one knows $q$, it is possible to calculate $T$. Morganti \etal\  (2001) calculate a value of $q= 10^{53}$ photons s$^{-1}$ sr$^{-1}$ to account for the H$\alpha$ luminosity in the filaments. With this incident photon flux, it is possible to raise the grains to 67 K. Since the dust temperature is such a weak function of the incident power, one would only require $\sim$0.01\% of power from the beamed flux in order to raise the dust temperature to $\sim$14 K. While this is plausible, the beamed flux would have to have a sufficient opening angle to illuminate both dust clouds entirely. The northern cloud is roughly 5 kpc across and at the distance of the cloud (15 kpc) this would require that the photon beam have an opening angle of at least $\sim$20\degree. This is uncomfortably large for a beamed source and we note that this minimum estimate does not engulf the optical filaments to the east. Also there is no known evidence of optical filaments to the south of Cen A, suggesting that the ionising beam can't be  interacting with the southern cloud at all.

\subsubsection{X-ray heating}
X-ray observations of Cen A have revealed that the galaxy has an X-ray jet and exists inside a diffuse X-ray halo (Kraft \etal, 2009; Turner \etal, 1997; Feigelson \etal, 1981). The model of Temi \etal\ (2003) takes into account the effects of a diffuse X-ray medium in which the dust grains are situated, and has been found to impact only small ($<0.1$ \micron) grains. It is possible that the presence of discrete X-ray sources might have a more significant effect. 

In a recent study, Kraft \etal\ (2009) reported the discovery of several bright X-ray knots on the eastern edge of the northern middle lobe, which places them to the north and east of the northern dust cloud. The nearest X-ray knot to the northern cloud (N4 in the notation of Kraft \etal) has a luminosity of only $\sim5 \times 10^{4} {\rm L}_{\odot}$ and therfore cannot be a potential heating source. 

A sufficiently bright low mass X-ray binary (LMSB) would also have the potential to heat the cloud. From Figs. 2 \& 3 in Kraft \etal\ there are two compact X-ray sources close to the location of the dust clouds. Using the XMM-Newton archive the two sources were located in the original data and found to have luminosities of L$_X$(north) $= 2.9 \times 10^{37}$ erg s$^{-1}$ and L$_X$(south) $= 3.6 \times 10^{36}$ erg s$^{-1}$. The luminosities are orders of magnitude below that required to power the dust clouds, so we conclude that X-ray emission either by diffuse hot gas or compact sources is not strong enough to account for the observed luminosity in the sub-mm.

\subsubsection{Jet mechanical heating}
The final situation to consider is a direct interaction between the radio jet and the dust cloud. In their comparison of the jet power to the X-ray power of the bright knots, Kraft \etal (2009) estimate that the jet itself has a mechanical power of $6 \times 10^{42}$ erg s$^{-1}$. Fig. \ref{fig7} indicates that the jet forms a cylinder from the NE inner lobe to the northern middle lobe. If it is assumed that the power is uniformly distributed, then it is possible to estimate the amount of mechanical power deposited into a single grain and hence deduce the dust grain temperature using Eqn. (\ref{eqn2}).

We approximate the jet as a cylinder $\sim$4\arcmin\ in diameter, which at the distance of Cen A is equivalent to $\sim$4.2 kpc. The cross-section of the cylinder at the distance of the northern dust cloud ($\sim$15 kpc) as seen from the nucleus of Cen A subtends an angle of $\sim$16\degree. Hence the jet subtends a solid angle of $\sim$0.005 sr at the distance of the northern dust cloud. This gives the jet a flux density of $\sim$1.2 $\times 10^{45} $erg s$^{-1}$ sr$^{-1}$. This equates to $\sim$3.6 $\times 10^{-10} $W per grain. Balancing this power with Eqn (\ref{eqn2}) yields a dust temperature of $\sim$2600 K. If all the mechanical energy were dumped into the dust, the dust would not survive, so the transfer of jet mechanical energy would have to be incredibly inefficient. The southern cloud also poses a problem for this mechanism, since there is no evidence of a jet-cloud interaction in the region surrounding the southern cloud. Despite the striking alignment between the dust clouds and the jet, the jet appears to have little influence on the dust, suggesting that the alignment is purely a projection effect.

We come to the same conclusion as Stickel \etal\ (2004). The simplest scenario that satisfactorily explains the origin of the dust is one in which the dust and gas in the clouds were once part of the late-type galaxy that merged with Cen A to produce its prominent dust lane. The only heating source that can account for the observed temperatures in both of the clouds is heating from the starlight of the evolved stellar population within Cen A.

\section{Conclusions}
\label{conclusion}
The results of these {\it Herschel}-SPIRE observations of Cen A have revealed two dust clouds external to Cen A and co-aligned with the axis of the AGN jet. The northern cloud lies at a projected distance of $\sim$15 kpc from the centre of Cen A and has been previously detected, while the southern cloud, at a distance of $\sim$12 kpc, is a new detection in the sub-mm. We have used SED analysis and a single-temperature, modified blackbody spectrum to fit the {\it Herschel}-SPIRE data and reprocessed MIPS data. The resulting fits show that the two clouds are thermal in origin and have roughly the same dust temperature and mass: T$_{north} = 12.6^{+1.1}_{-1.2}$ K, T$_{south} = 15.1^{+1.7}_{-1.6}$ K; $\rm{log} (M_{north} / M_{\odot}) = 5.8^{+0.2}_{-0.2}$, $\rm{log} (M_{south} / M_{\odot}) = 5.6^{+0.2}_{-0.2}$. The measured values for the northern cloud are consistent with previous measurements based on ISO data (Stickel \etal, 2004).

The clouds reside at the termini of the partial \hi\ ring that surrounds Cen A where the \hi\ column density is highest. The measured gas-dust ratios and \hi\ / H$_2$ ratios are typical of a late-type gas-rich galaxy, which support the hypothesis that the dust and gas originated in the galaxy which has since merged with Cen A to form its dusty disk. Dust was not unambiguously detected in other parts of the \hi\ ring and we attribute this to the dust having a column density below the {\it SPIRE} detection threshold rather than having a lower temperature or an absence of dust altogether.

The dust clouds have also been detected in \hi, CO, 24 \micron\ and the southern cloud, at least, also exhibits 8 \micron\ PAH emission. SFRs have been calculated based on UV-FIR tracers but the SFRs are too low to have produced the inferred mass of dust in a reasonable time frame. The low SFRs and the lack of UV emission in the southern cloud imply that the dust emission is unrelated to star formation. Dust heating by the evolved stellar population is the simplest mechanism capable of explaining the observed temperatures for both clouds. The jet therefore is unlikely to play a major r\^ole in the heating process and its alignment with the clouds is purely coincidental.

\section*{Acknowledgments}
{\it SPIRE} has been developed by a consortium of institutes led by Cardiff University (UK) and including Univ. Lethbridge (Canada); NAOC (China); CEA, LAM (France); IFSI, Univ. Padua (Italy); IAC (Spain); Stockholm Observatory (Sweden); Imperial College London, RAL, UCL-MSSL, UKATC, Univ. Sussex (UK); and Caltech, JPL, NHSC, Univ. Colorado (USA). This development has been supported by national funding agencies: CSA (Canada); NAOC (China); CEA, CNES, CNRS (France); ASI (Italy); MCINN (Spain); SNSB (Sweden); STFC (UK); and NASA (USA). HIPE is a joint development (are joint developments) by the Herschel Science Ground Segment Consortium, consisting of ESA, the NASA Herschel Science Center, and the HIFI, PACS and SPIRE consortia.

\bibliographystyle{mn2e}
\bibliography{auld-mnras-2011}
\nocite{1}
\nocite{2}
\nocite{3}
\nocite{4}
\nocite{5}
\nocite{6}
\nocite{7}
\nocite{8}
\nocite{9}
\nocite{10}
\nocite{12}
\nocite{13}
\nocite{14}
\nocite{15}
\nocite{16}
\nocite{17}
\nocite{18}
\nocite{19}
\nocite{20}
\nocite{21}
\nocite{22}
\nocite{23}
\nocite{24}
\nocite{25}
\nocite{26}
\nocite{27}
\nocite{28}
\nocite{29}
\nocite{30}
\nocite{32}
\nocite{33}
\nocite{34}
\nocite{35}
\nocite{36}
\nocite{37}
\nocite{38}
\nocite{39}
\nocite{40}
\nocite{41}
\nocite{42}
\nocite{43}
\nocite{44}
\nocite{45}
\nocite{46}
\nocite{48}
\nocite{49}
\nocite{50}
\nocite{51}
\nocite{52}
\nocite{53}
\nocite{54}
\nocite{55}
\nocite{56}
\nocite{57}
\nocite{58}
\nocite{59}
\nocite{60}
\nocite{61}
\nocite{62}
\nocite{63}
\nocite{64}
\nocite{65}
\nocite{66}
\nocite{67}
\nocite{68}
\nocite{69}
\nocite{70}
\nocite{71}
\nocite{72}

\end{document}